\documentclass[prl,twocolumn,showpacs]{revtex4}  % for review and submission
\usepackage{graphicx}  % needed for figures
\begin{document}

% the following line is for submission, including submission to the arXiv!!
%\hspace{5.2in} \mbox{Fermilab-Pub-04/xxx-E}

\title{On the Mechanism of the Volume Reflection
of Relativistic Particles \footnote{ JETP Letters, 2008, Vol. 87, No. 7, pp. 349 -353} }

%\input author_list.tex       % D0 authors (remove the first 3 lines
                             % of this file prior to submission, they
                             % contain a time stamp for the authorlist)
                             % (includes institutions and visitors)
\author{G.\,V.\,Kovalev \/\thanks
{e-mail: kovalevgennady@qwest.net}}

%%% author(s) - for colontitle (at the top of the page)
%\rauthor{G.\,V.\,Kovalev}

%%% author(s) - for table of contents
%\sodauthor{Kovalev}

%%% author's address(es)
\affiliation{North Saint Paul,  MN 55109, USA }

%%% dates of submition & resubmition (if submitted once, second argument is *)
%\dates{12 November 2012}

\date{Feb. 15, 2008}

\begin{abstract}
The mechanisms of the volume reflection of positively and negatively charged relativistic particles in a bent crystal have been analyzed. It has been shown that the empty core effect is significant for the negatively charged particles. The average reflection angle of the negatively charged particles has been determined and the conditions for the observation of the reflection and refraction are discussed.
\end{abstract}

%%% PACS numbers
%\PACS{03.65.Ge; 03.65.Fd; 03.65.׷; 03.65.Db;  02.30.Em}
\maketitle

%\section{\label{sec:level1}First-level heading}
% sections are not used for PRL papers
The experiments reported in [1-3] confirmed the
effect of the volume reflection of 1-, 70-, and 400-GeV
protons in a bent Si crystal, which was revealed by
Taratin and Vorobiev [4, 5] using the Monte Carlo
method, and demonstrated the possibility of its application to collimate accelerated beams [6]. The numerical
calculations performed in [4, 5, 7] also indicated the
volume reflection of negatively charged particles, but
the corresponding reflection angle is smaller than that
for positively charged particles. However, the reflection
of the negatively charged particles differs in nature
from the reflection of the positively charged particles
and requires a more detailed analysis. Indeed, within
the framework of classical mechanics, the reflection
and scattering of the negatively charged particles in the
field of a one-dimensional potential well (see Fig. 1a)
are absent, whereas the grazing incidence of the postively charged particles on a one-dimensional barrier,
  $\alpha < \theta_L$ ($\alpha$
 is the angle between the particle momentum
and the boundary of the one-dimensional barrier and 
 $\theta_L$ is the Lindhard critical angle, see Fig. 1b), is accompanied by the complete reflection. For the case of a centrally symmetric potential, this phenomenon is responsible for the volume reflection of positively charged relativistic particles in a bent crystal. Indeed, the impact parameter  $b$  measured from the tangential edge (point  $T$
in Fig. 1c) of the centrally symmetric ring barrier with
the height   $U_o$   is given by the expression 
 $b=R(1-cos(\alpha))$  and the average reflection angle is specified by
the integral
\begin{eqnarray}
\bar{\chi}=\frac{1}{b_{max}} \int_{0}^{b_{max}}2\alpha \; d b ,
\label{avarage_angleIntPos}
\end{eqnarray}
Here, the maximum impact parameter, $b_{max} =R(1-cos(\theta_L))$,  depends on the critical channeling angle $\theta_L=\sqrt{2U_{o}E/(p^2c^2)}$ 
 and the potential-barrier radius $R$.   For small angles  $\alpha,\theta_L <<1$ ,
  $b_{max}=R\theta_{L}^{2}/2 $,   $d b=R \alpha  \; d\alpha$, and  Eq. (1) provides the average reflection angle
\begin{eqnarray}
\bar{\chi}=4\theta_L/3,
\label{avarage_anglePos}
\end{eqnarray}
coinciding with the estimate obtained for the bent crystal {Eq. (18) in [8]}.

On the contrary, it is well known that the reflection
is absent in the case of scattering on the centrally symmetric potential well with the depth  $-U_o$.   In this case,
only refraction occurs for all impact parameter values
(see problem 2 in Sect. 19 in [9]). Thus, the reflection
of the negatively charged particles in the numerical
simulation is at first sight surprising. Nevertheless, as
shown in [8], the negatively charged particles scattered
on a ring potential undergo reflection. Since the presence of the inner potential wall distinguishes the ring
potential from the potential well, it is natural to think
that this reflection is the reflection from the inner potential wall of the ring potential similar to the reflection
from the outer wall of the potential barrier. However, as
shown below, the determining factor of the reflection of
the negatively charged particles in a crystal is the effect
of a specific deflection called the empty core effect [8].

\begin{figure}
	\centering
		\includegraphics{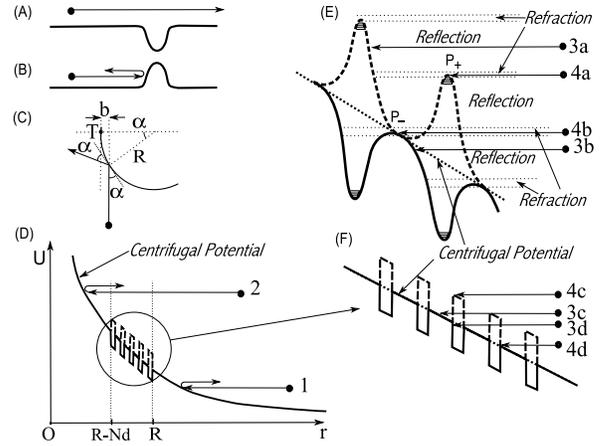}
	\caption{One-dimensional motion of the (a) negatively and
(b) positively charged particles, (c) the reflection of a positively charged particle from the centrally symmetric barrier,
(d) the effective potential of the system of rectangular ring
potentials, (e) the reflection and refraction regions, and (f)
the liner approximation for the effective potential.}
	\label{fig:CentrifugalPotential}
\end{figure}

For the scattering in a central field (and in a uniformly bent crystal), the following effective potential is
introduced:
\begin{eqnarray}
U_{eff}(r)=U(r)+ \frac{p_{\infty}^2c^2b^2}{ 2E r^2},
\label{Effective_Potential}
\end{eqnarray}
where $b$ is the impact parameter measured from the
center and the remaining symbols are the same as in [8].
In this potential, any infinite motion formally looks like
a “reflection;” i.e., a particle comes from infinity, stops
at the turning point, and goes away to infinity.

We emphasize that such treatment of the reflection
is not necessarily valid [3]. The first example is trivial:
particle $1$  in Fig. 1d is “reflected” at the distance  $b >R$  
from the center of the effective potential, which
includes only the centrifugal term, because $U(r) = 0$ in
this region. This motion in the Cartesian coordinates is
a rectilinear motion. For the second example, particle $2$ 
in Fig. 1d is “reflected” at the distance $b < R  – N d$ from
the center of the ring potential, where also $U(r) = 0$.
This is the empty core effect. As shown in [8], two
opposite situations are possible. First, a particle with
any trajectory similar to trajectory $2$ in Fig. 1d is
reflected (has the angle $\chi >0$) in an arbitrary negative
nonsingular potential localized in a ring. Second, a particle whose trajectory is similar to trajectory $2$  in Fig. 1d
is refracted (has the angle $\chi < 0$) in a positive nonsingular ring potential. In both cases, the turning point is the same and lies on the centrifugal curve.
Finally, let us consider the examples where the turning point lies in the crystal potential $U(r)$.   They are in
close connection with the experiments on the reflection
of relativistic particles. Figures 1d and 1f show the
effective potential of the system of periodic rings used
in [8]. The real crystal potential is a smooth function
and, if the Tsyganov criterion $R > \frac{p_{\infty}^2c^2}{U(r)^{'}}$ [10] is
satisfied, the effective potential has the system of
smooth maxima and minima corresponding to each
bent potential plane (see Fig. 1e).

In the 1950s, Ford and Wheeler [11, 12] showed that, if the effective potential for a certain impact parameter has a smooth local maximum, the deflection
function [13]
\begin{eqnarray}
\chi(b)=\pi-2\sqrt{\frac{p_{\infty}^2c^2b^2}{ 2E}}\int^{\infty}_{r_{o}}\frac{\frac{1}{r^2}d r }{ \sqrt{\frac{p_{\infty}^2c^2}{ 2E}-U(r)-\frac{p_{\infty}^2c^2b^2}{ 2E r^2}}},
\label{deflection_function}
\end{eqnarray}

has a negative logarithmic singularity; i.e., the particle
incident on the maximum of the potential undergoes an
indefinite number of rotations about the center. This
phenomenon of spiral or orbital scattering was well
known in low-energy atomic collisions [14, 13]. Under
certain conditions, this mechanism also occurs for relativistic particles. If the particle trajectories touch the effective potential at the local-maximum points (trajectories 4a and 4b in Fig. 1e), not only the radial velocity
of the particle,  $v_r$, but also the radial force  $-\partial U_{eff}/\partial r$ 
(and, therefore, the radial acceleration   $\dot{v}_r$ ), vanish at
these points. This means that the particle motion in the
radial direction ceases, whereas the particle continues
to move in the tangential direction (the circle with the
radius $r_0$ is a limit cycle which the particle approaches
along a spiral). This is the classical spiral-scattering
mechanism. If losses were absent, the particle could
execute an infinite number of rotations about the center.
This position is evidently unstable and small fluctuations move the particle out of this state. Resonant scattering is a quantum-mechanical analog of this phenomenon [15].  Strictly speaking, only one trajectory satisfies the spiral scattering conditions. For trajectories
close to this trajectory, this effect in classical mechanics
is manifested as the refraction: the particle, passing a
small part of the angular path towards the potential
bend, is deflected and moves away from the center
along the symmetric asymptotic curve. The reflection
and refraction regions are shown in Fig. 1e by the
dashed lines.

However, in the experiments on the reflection of
positively charged particles [1-3], the volume capture
of protons into the channeling regime was observed
instead of the spiral scattering. In my opinion, this is
due to two important causes. First, the crystal nucleus
density and electron density (dark regions in Fig. 1e)
are high at maxima for positively charged particles.
Second, the spiral scattering and refraction of the positively charged particles ensure a quite long-term presence of particles near the maxima of the effective
potential and, therefore, strongly increase the intensity
of the volume capture. The maximum of the potential
for negatively charged particles is a wider concave
parabola; hence, the refraction region for the negatively
charged particles should be wider. The crystal nucleus
density and electron density almost vanish at the maxima for the negatively charged particles. Therefore, the
stability of the refraction trajectories for the negatively
charged particles in a narrow impact-parameter range
should be much higher. As a result, a fraction of the
negatively charged particles in the narrow impact-
parameter range can move towards the crystal bend
without the mechanism of volume capture into the
channeling regime due to the refraction near the local
maxima of the potential and a related mechanism, that
is, the spiral scattering. In this case, the peak characteristic of the volume capture of the positively charged
particles should be absent for the capture of the negatively charged particles. Thus, the effect of the refraction and spiral scattering can be noticeable in a detailed
comparison of the scatterings of positively and negatively charged relativistic particles in the region traditionally treated as the volume capture region.
The exact solution of the problem of scattering on
the potential of the periodic system of rectangular rings
was considered in [8], where the average angle of the
reflection of positively charged relativistic particles that
describes the experiments reported in [1-3] was also
estimated. In the model of the rectangular rings, the
potential is not a smooth function and the radial force
does not vanish at the tangent points (see, e.g., lines 4c
and 4d in Fig. 1f); hence, the spiral scattering is absent
in this model. However, it is applicable for describing
the reflection (see lines 3c and 3d in Fig. 1f) including
the reflection of negatively charged relativistic particles; the details of this description are given below.
First, most trajectories of negatively charged particles have the turning points lying on the centrifugal
potential curve between potential wells (see curve 3c in
Fig. 1f) and a few trajectories enter the potential well
regions (see curve 3d in Fig. 1f). Since the crystal
potential lying above the trajectory (line 3c) does not
affect the behavior of the particle, it is easy to see that
the scattering of the particle with a trajectory similar to
line 3c has the same nature as the scattering of the particle with trajectory 2 in Fig. 1d; i.e., the turning angle,
in this case, is determined by the empty core effect.
Since the potential lying under the particle trajectory is
negative, the reflection occurs for trajectories of type
3c. This qualitatively explains Figs. 2c and 2d in [8] and
Fig. 2a in this paper, where the angles in the region on
the left of the point 1 –  are positive. Formula (15) in
[8] contains a misprint. The correct expression for the
scattering of negatively charged particles on one ring
that was used to plot Figs. 2c and 2d in [8] has the form

\begin{equation}
\alpha(\hat{b})_{-}=
\label{DeflectionFunction_Ring_v2}
\end{equation}
\begin{eqnarray}
\left \{
\begin{array}{ll}
(\sqrt{1-\hat{b}^2}-\sqrt{\Phi-\hat{b}^2})-(\sqrt{1-\hat{b}_{a}^2}-\sqrt{\Phi-\hat{b}_{a}^2}),& \\  for \; \; 0 < \hat{b} <(1-\hat{a});\\  
\sqrt{1-\hat{b}^2}-\sqrt{\Phi-\hat{b}^2}+\sqrt{\Phi-\hat{b}_{a}^2},\nonumber\\for \; \;(1-\hat{a}) < \hat{b} <(1-\hat{a})\sqrt{\Phi};\nonumber\\ 
\sqrt{1-\hat{b}^2}-\sqrt{\Phi-\hat{b}^2},\; \;for\; \; (1-\hat{a})\sqrt{\Phi} < \hat{b}< 1;\\
0, \; \;for \; \;1 < \hat{b}.
\end{array} \right.
\end{eqnarray}

Some of Eqs. (16) in [8] for the maximum and minimum deflection angles of negatively charged particles should be written as
\begin{eqnarray}	
\alpha_{max -} =\sqrt{2\hat{a}}- \sqrt{2\hat{a}-\phi_o}-\sqrt{-\phi_o}& &\nonumber\\
\alpha_{min -}=- \sqrt{-\phi_o}.
\label{MaxMin}
\end{eqnarray}

They are obtained from Eq. (5) using the substitutions
  $\hat{b}=(1-\hat{a})$ and $\hat{b}=1$ and the smallness conditions
 $\hat{a}<<1$ and $\hat{a}^2 <<\hat{a}$. When the well is sufficiently deep,  $|\phi_o| > 2\hat{a}$,  these formulas are modified to the form
\begin{eqnarray}	
\alpha_{max -} =\sqrt{2\hat{a}}- \frac{\hat{a}}{\sqrt{|\phi_o|}}& &\nonumber\\
\alpha_{min -}=- \frac{\hat{a}}{\sqrt{|\phi_o|}}.
\label{MaxMinDeep}
\end{eqnarray}

In this case, the range $(1-\hat{a})\sqrt{\Phi} < \hat{b}< 1$ contracts to
zero. This is seen in Figs. 2d and 3d in [8] (cf. Figs. 2c
and 3c in [8]), where the lower part of the deflection
function corresponding to the bottom of the potential
well disappears.

The general expression for the deflection function of
negatively charged particles in the system of rectangular rings [8], when the impact parameter  lies in the
ring   $ (1-(k+1)\hat{d} ) < \hat{b} <(1-k\hat{d})$,  has the form
\begin{equation}
\chi(\hat{b})_{-}= 2 \sum_{i=0}^{k-1} \alpha_i(\hat{b}) + 
\label{DeflectionFunction_SystemN}
\end{equation}
\begin{eqnarray}
+2\left \{
\begin{array}{ll}
(\sqrt{1-\hat{b}_{k}^2}-\sqrt{\Phi-\hat{b}_{k}^2})-(\sqrt{1-\hat{b}_{ka}^2}-\sqrt{\Phi-\hat{b}_{ka}^2}),& \\  for \; \; (1-(k+1)\hat{d} ) < \hat{b} <(1-\hat{a}-k\hat{d} );\\  
\sqrt{1-\hat{b}_{k}^2}-\sqrt{\Phi-\hat{b}_{k}^2}+\sqrt{\Phi-\hat{b}_{ka}^2},\nonumber\\for \; \;(1-\hat{a}-k\hat{d}) < \hat{b} <(1-\hat{a}-k\hat{d})\sqrt{\Phi};\nonumber\\ 
\sqrt{1-\hat{b}_{k}^2}-\sqrt{\Phi-\hat{b}_{k}^2},\; \;\nonumber\\for\; \; (1-\hat{a}-k\hat{d})\sqrt{\Phi} < \hat{b}< (1-k\hat{d}).\nonumber\\
\end{array} \right.
\end{eqnarray}

The deflection function calculated using Eq. (8) with
various values of the parameter $ {2U_o E }/(p_{\infty}^2c^2\hat{a})$
 is shown in Fig. 2a as a function of the potential depth $U_o$.

\begin{figure}
	\centering
		\includegraphics{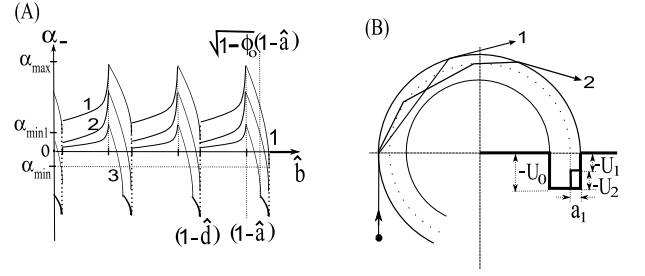}
	\caption{ (a) Deviation functions for negatively charged particles at $\phi_o/\hat{a}$:  1 - $\phi_o/\hat{a}=12$,    2 - $\phi_o/\hat{a}=4$, 3 -  $\phi_o/\hat{a}=1.2$;   (b) The tangential
scattering trajectory on (1) the square potential well and (2)
step potential with the same maximum depth as for the
square potential well $U_0 = U_1 + U_2$.}
	\label{fig:DeflectionFunctionSN}
\end{figure}

The average reflection angle for negatively charged
particles can be calculated from the potential period
nearest to the edge. When the reflection from the inner
potential wall is disregarded, this angle is determined
by the integral 
\begin{eqnarray}	
\bar{\alpha}_{-} = \frac{1}{\hat{d}-\hat{a}} \int^{1-\hat{a}}_{1-\hat{d}} 
\Bigl((\sqrt{1-\hat{b}^2}-\sqrt{\Phi-\hat{b}^2})-\nonumber\\
(\sqrt{1-\hat{b}^2_{a}}-\sqrt{\Phi-\hat{b}^2_{a}})\Bigl) d\hat{b} 
\label{AverageAngleNReflect}
\end{eqnarray}
with subsequent expansion in small quantities  $\hat{a},\hat{d}$
and  $\phi_o$. A s a result, the expression for the average reflec
tion angle $\chi_{-}=2\cdot\bar{\alpha}_{-}$
 for negatively charged particles is
obtained in the form
\begin{eqnarray}	
\bar{\chi}_{-} = \frac{2}{3(\hat{d}-\hat{a})}\Bigl((2\hat{d}-2\hat{a}-\phi_o)^{3/2}+(2\hat{d})^{3/2}+\nonumber\\ (2\hat{a}-\phi_o)^{3/2} - (-\phi_o)^{3/2}- (2\hat{d}-\phi_o)^{3/2}\nonumber\\-(2\hat{a})^{3/2}-(2\hat{d}-2\hat{a})^{3/2}\Bigl).
\label{AverageAngleNReflectF}
\end{eqnarray}

Formula (10) applied to possible experiments on the
scattering of 1-, 70-, and 400-GeV antiprotons with the
same geometry of crystals as in [1-3] yields the average
reflection angles 14.8, 4.3, and 1.8 mrad, respectively.

In conclusion, let us discuss the possibility of an
improvement in the rectangular-ring model that could
include the spiral scattering. The refraction angle of the
tangential trajectory for the system of rectangular steps
that is close to the actual potential curve is larger than
the angle for the rectangular well. This relation is seen
from the following example. For the tangential trajectory of the negatively charged particle, the refraction angle [minimum angle in Eqs. (6)] for the square potential well with the depth $U_0$ is $\alpha_{min -}=- \sqrt{-\phi_o}$, where
 $\phi_{o}=-\frac{2U_{o} E}{p_{\infty}^2 c^2}$.  

Let us consider a step potential shown in Fig. 2b with arbitrary $U_1$ and $U_2$ values satisfying the condition $U_0 = U_1 + U_2$. For the same tangential trajectory, the deflection angle on the step with the depth $-U_1$ is   $- \sqrt{-\phi_1}$,  where $\phi_1=-\frac{2U_{1} E}{p_{\infty}^2 c^2}$.   

If the step width $a_1$ is chosen so that the trajectory touches the boundary of the step or intersects the boundary (such choice is always possible), the secondary refraction
appears at the angle $-\sqrt{\phi_2}$,  where  $\phi_2=-\frac{2U_{2} E}{p_{\infty}^2 c^2}$.

The total refraction angle is  $-(\sqrt{-\phi_1}+\sqrt{-\phi_2})$ with the
condition $\phi_0 = \phi_1 + \phi_2$.   Since the inequality $\sqrt{A}+\sqrt{B}>\sqrt{A+B}$
 is always valid, the total scattering angle on the step potential,  $\sqrt{-\phi_1}+\sqrt{-\phi_2}$, is larger than the angle $\sqrt{-\phi_1-\phi_2}=\sqrt{-\phi_o}$
 on the square well potential, although the maximum depths of these potentials are
the same. Subsequently dividing the potential into smaller steps, one can unlimitedly increase the refraction angle, thus realizing the spiral scattering on the step potential of a finite depth.

%\bibliographystyle{jetpl}
%\bibliography{../../Focusing_and_Channeling_in_Crystals/chan02}

\end{document}